# Power law behavior associated with a Fibonacci-Lucas model and generalized statistical models

Aram Z. Mekjian

Department of Physics and Astronomy, Rutgers University, Piscataway, NJ 08854

Abstract

A Fibonacci-Lucas based statistical model and several other related models are studied. The canonical and grand canonical partition functions for these models are developed. Partition structure such as the distribution of sizes as in a cluster distribution is explored. Ensemble averaging over all partitions leads to a scale invariant power law behavior at a particular critical like point. The canonical ensemble of the Fibonacci-Lucas case involves the Gegenbauer polynomial. The model has a hyperbolic power law behavior, a feature linked to the golden mean ratio of two adjacent Fibonacci numbers and also the connection of Lucas numbers to the golden mean. The relation to other power law behavior, such as Zipf's and Pareto's law, is mentioned. For the cases considered, the grand canonical ensemble involves the Gauss hypergeometric function $_2F_1[a,b,c,z]$ with specific values for $a,b,c$. The general case has a variable power law behavior with $\tau$ exponent equal to $1+c-a-b$. An application with $a=1/2, b=1, c=3$ and thus $\tau=5/2$ very closely approximates Bose-Einstein condensation. The zeta function $\varsigma(3/2)=2.61$ of the exact theory is replaced with 8/3 and $\varsigma(5/2)=1.34$ with 4/3. At the condensation point the number of cycles of length $k$ falls as a scale invariant power law. The cycles, which arise from permutation symmetries associated with Bose-Einstein statistics, can be viewed as links in a complex network of connections between particles. This scale invariant power law parallels that seen in complex networks. The growth of the network is developed using recurrence properties of the model. Constraints imposed by the canonical ensemble and associates correlations lead to some number theoretic connections between Fibonacci and Lucas numbers as an incidental consequence of this approach.

I. Introduction

Clustering and fragmentation phenomena, cascading or branching processes, partition theory, connectivity, networking and percolation problems are of current interest in many fields. As a specific example from physics is a cluster distribution which gives the number of clusters of a specific size as a function of the number of constituents that makes up the cluster. A cluster model will list all the ways an initial set of objects can be partitioned into subgroups of all sizes consistent with initial constraints. Clustering into groups can also be viewed as a connectivity problem as in percolation. Multifragmentation is a phenomena that occurs in heavy ion collisions where the colliding nuclei break into subgroups or pieces from the original nuclei. A similar division into groups occurs in Bose-Einstein and Fermi-Dirac distributions which involve permutations and their cycle class decomposition [1,2,3]. In this case a permutation is specified by listing the number of cycles with associated cycle lengths that make up the permutation. A unit cycle occurs when no interchange has happen and a cycle of length two occurs when two particles or objects are interchanged, etc. This cycle class representation can also be used to discuss Bose-Einstein condensation, a topic of great current interest. Feynman also used the cycle class picture as the basis for an investigation of the superfluid $\lambda$ transition. In mathematics, partitions appears in the theory of integer partitions which list all the ways an integer can be broken up into groups whose sum is the original integer. Discussions of integer partitions can be found in ref.[4,5]. In mathematical biology, partitions are introduced when discussing diversity. Diversity, for example, is measured by looking at the number of copies a given gene has in a sampling..

Size distributions appear in many diverse fields which, to list several cases, are: 1. percolation clusters in percolation studies on various lattice types [6], 2. Ising clusters in studies of the statistical mechanics of the Ising model [7], 3. in the size distribution of meteorites [8], 4. in sandpile slides and models of self organized criticality[9,10] and 5. avalanche slides [11], 6. in the number of earthquakes as a function of their energy [12], 7. in the spectral decomposition of noise as a function of frequency with its interesting low frequency 1/f flicker behavior. Ref[13] discusses $1/f$ noise and other distribution with long power law tails. Also connectivity of networks is of recent interest stemming from the work in several papers [14,15,16]..

One of the remarkable features of different types of distributions is the appearance of a power law behavior. In cluster yields the number of clusters $n_k$ of size $k$ can fall as a power law, or $n_k \sim 1/k^\tau$ with $\tau$ an exponent. In physical phenomena, this happens at a critical point such as in the Ising model or a self-organized critical point such as in sandpile models. In percolation models a power law distribution of percolation clusters arises at the point at which an "infinite" cluster appears and these coupled features happen at a specific value of the bond or site probability. In linguistics a power law behavior is present where it is known as Zipf's law [17]. The power law is seen in the distribution of words in a book when each word is ordered according to its frequency of occurrence. In economic phenomena, Pareto's law[18] is an observed power law behavior in the distribution function of incomes plotted against income amount. In social phenomena, Lotka's law [19] is an observed power law relation between the number of authors publishing $n$ papers when plotted against $n$. More recently, power laws associated with small world networks have been noted [14,15,16]. Extensions of the classic Erdos-Renyi graph theory[20] have been a developed for such studies[14-16,21-23]. A recent summary of power laws can be found ref.[24,25]. Ref.[26] contains a discussion of why power laws are present from gene families, genera, protein family frequencies, income and internet file sizes.

The importance of powers laws and its relation to self-similarity and also fractal behavior was initially stressed by Mandelbrot [27] and later by Schroeder [28]. The backbone of fractals is iteration and the archetype example of iteration is the generation of Fibonacci numbers. The golden mean $\phi = (1+\sqrt{5})/2$ is the limiting ratio of two adjacent Fibonacci numbers. Fibonacci numbers, given by the sequence 1,1,2,3,5,8,13,…and Lucas numbers, given by the sequence, 2,1,3,4,7,11,….have the property that each term in the sequence is generated by summing the two previous numbers starting with the third term. The Fibonacci sequence has been a prototype model for growth which in this case is unbound. Generators for the Fibonacci sequence also play an important role in understanding features associated with these numbers. Many textbooks exist that summarize properties of Fibonacci and Lucas numbers [29,30] and their applications and ubiquitous presence. Because of their connection with the Golden mean, the Fibonacci and Lucas numbers appear in theories which try to give a mathematical basis for aesthetics [31]. Weyl noted the importance of the Fibonacci numbers in his book on symmetry [32]. Schroeder [28] has discussed the appearance of the Fibonacci numbers in several examples such as in Arnol'd's cat map which has hyperbolic fixed points. As noted in this reference maps with hyperbolic fixed points, {maps which expand in one direction and contract in the orthogonal direction} are the hallmark of chaotic systems in energy preserving Hamiltonian systems. Fibonacci models appear in population biology [33].

Since Fibonacci and Lucas numbers have played such an important role in so many areas, it therefore seemed worthwhile to explore their usefulness and properties when used as a weight function in the theory of partition structures. The weight function is used to generate the canonical ensemble by summing over all possible partitions. The Fibonacci/Lucas model will be shown to give rise to an analytic canonical partition function which is a Gegenbauer polynomial. The associated grand canonical partition function can be written in terms of Gauss hypergeometric functions. As will be shown, such a weight gives rise to a hyperbolic power law. Hyperbolic power law behavior are associated with Zipf's and Pareto laws. Simple cycle class permutation weights also have a pure hyperbolic power law dependence as mentioned in sec.II.B. The Fibonacci/Lucas model will be shown to be a convolution of two permutation models involving either of the two Golden mean numbers $\phi$ or $\phi'$ with $\phi' = (\sqrt{5}-1)/2 = 2/(1+\sqrt{5}) = 1/\phi$.

This paper is organized as follows. The next section, sect.II contains the main theoretical framework for this paper and it is divided into various subsections II.A-II.F. These subsections are as follows. The first subsection contains a discussion of partitions and partition weights. The theory of partitions is used to illustrate how a group of objects can be divided into subgroups of varying sizes. Partition weights are introduced using two examples taken from combinatorial analysis. A brief summary of combinatorial analysis can be found in chapter 24 in Abramowitz and Stegun [34]. Based on these two examples, the partition weight is generalized. Several specific and different applications are listed in tables to illustrate the wide spectrum of results that can be studied using the generalized scheme. This subsection is then followed by four subsections, II.B – II.E where four specific examples are discussed. The first three examples have already been considered in previous papers [35-38] and are therefore only briefly summarized. They are included because results obtained from them can be linked to results from the Fibonacci/Lucas weight. These cases also have common features which are important for developing a generalized picture. Example1 appears in the theory of permutations which is also important in Bose-Einstein and Fermi-Dirac gases and Bose-Einstein condensation which is developed further in subsection II.F. Example 2 appears in quantum optics [39] and is used as an example of an application to photon count distributions. Example 2 is originally due to Glauber [40] and has been used in particle physics. This distribution also has its counterpart in a Feynman-Wilson gas [36,41]. This second example involves the Catalan number and a branching model based on it is mentioned. The first two examples are special cases of the third example which is related to a Levy distribution and contains a Levy index. The fourth example involves the Fibonacci/Lucas weight and is dealt with in greater detail. A generalized model based on a Gauss hypergeometric function is discussed in section II.F. The Gauss hypergeometric model has a variable power law exponent $\tau = 1 + c - a - b$ where $a, b, c$ appear in the hypergeometric function. This subsection also contains an approximate model for Bose-Einstein condensation. An application to networks is explored. The last main section, sect. III, contains the conclusions and summarizes the main findings.

II. Theoretical Framework
II.A Partitions and weights over partitions.

A partition of $A$ objects is a division of $A$ into subgroups of varying sizes. The partition can be specified by a set or vector $\vec{n} = (n_1, n_2, n_3, ..., n_A)$ or by an alternative notation $1^{n_1}, 2^{n_2}, 3^{n_3}, ..., A^{n_A}$. If $n_j = 1$, then $j^{n_j} = j$ for its notation in the partition. The index $k$ is the size of the group and $n_k$ is the number of groups of size $k$. A partition can be pictured as a block diagram as shown in Fig.1.

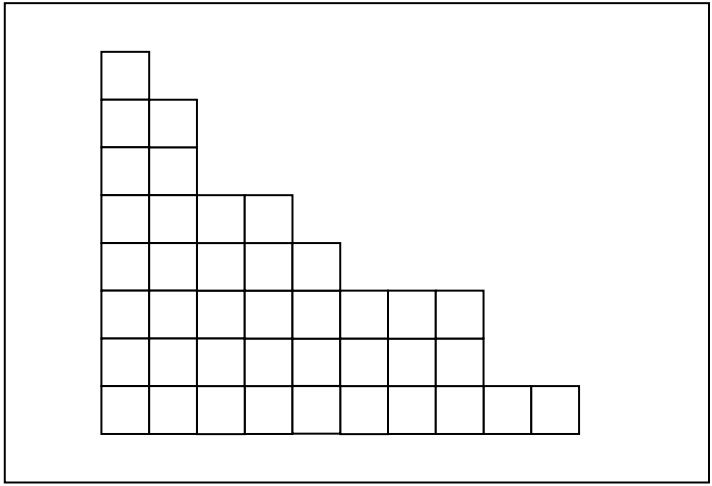

Fig1. A block diagram of a partition. The vertical number of boxes or blocks in a given column is $k$ and the number of columns with k boxes or blocks is $n_k$. The partition shown is $1^2, 3^3, 4, 5^2, 7, 8$. The multiplicity $m = \Sigma_k n_k$ is the number of columns which, in this case, is 10. The $A = \Sigma_k k n_k = 40$ is the total number of blocks. The total number of ways the blocks can be stacked is the total number of partitions. The extreme partitions are one column having all the blocks which has $m = 1$ and $n_A = 1$ or one row one block high which then has $m = A$ and $n_1 = A$. Table 1 lists various examples of partitions.

Table 1. Examples of partitions.

| Area | meaning of $k$ | meaning of $n_k$ |
| --- | --- | --- |
| Cluster or group structure | size of the cluster or group | number of clusters or groups of size $k$ |
| Permutations, Bose-Einstein & Fermi-Dirac problems | cycle length | number of cycles of length $k$ |
| Integer partitions | integer $k$ | number of times $k$ appears |
| Gene diversity | number of copies of an allele in a sample | number of different alleles each occurring $k$ times |
| Specie diversity | frequency of occurrence of a given specie | number of different species each appearing with frequency $k$ |
| Networks | size of a connection | number of connections of size $k$ |

In integer partitions, the $n_k$ is the number of times the integer $k$ appears in the decomposition of $A$. For example $A = 5$ has 7 separate partitions which are 5, 4+1, 2+3, 1+1+3, 1+2+2, 1+1+1+2, and 1+1+1+1+1. For the specific partitions 1+1+3, the associated $\vec{n} = (2,0,1,0,0)$ or, in the alternative notation, 1+1+3 is represented by $1^2, 3$ with the terms with $n_k = 0$ being suppressed in this notation. A constraint appears which is $A = \Sigma_k k n_k$. The multiplicity of a given partition $\vec{n} = (n_1, n_2, ..., n_A)$ is the sum of $n_k$: $m = \Sigma_k n_k$. The multiplicity $m$ ranges from $m = 1$ (for the partition of $A$ with $n_A = 1$ and all other $n_k = 0$) to $m = A$ (for the partition which has $n_1 = A$ and all other $n_k = 0$). The number of partitions $p(A)$ grows exponentially fast with $A$. For large $A$, $p(A) \sim e^{\pi\sqrt{2/3}\sqrt{A}} / 4\sqrt{3}A$ which is a result due to Hardy and Ramanujan. The number of partitions of $A$ with multiplicity $m$ is labeled $P(A, m)$. The $P(A, m)$ satisfy the recurrence relation: $P(A, m) = P(A-1, A-1) + P(A-m, m)$. For large $A$ and $m$ with $m \sim O(A^{1/3})$ $P(A, m) \approx A^{m-1} / (m!(m-1)!)$

Such decompositions appear in combinatorial analysis. One particular example is in the cycle class decomposition of a permutation. In this case the $n_k$ is the number of cycles of length $k$. The number of permutations of $A$ is $A!$. Of the $A!$ permutations, the number of permutations belonging to a particular $\vec{n} = (n_1, n_2, n_3, ..., n_A)$ or $1^{n_1}, 2^{n_2}, 3^{n_3}, ..., A^{n_A}$ is given by the Cauchy result [34]

$$M_2(\vec{n}, A) = \frac{A!}{1^{n_1} n_1! 2^{n_2} n_2! 3^{n_3} n_3!,..., A^{n_A} n_A!} \tag{1}$$

A factor $i^{n_i}$ appears because a given cycle in a permutation can be started with any member of the cycle. The $M_2(\vec{n}, A)$ satisfies the sum rule $A! = \Sigma_{\vec{n}} M_2(\vec{n}, A)$ over all partitions of $A$. The probability of a particular cycle class is $P(\vec{n}, A) = M_2(\vec{n}, A)/A!$.

A second example of a counting factor that appears in ref.[34] is the number of ways of partitioning a set of $A = n_1 + 2n_2 + ....An_A$ distinguishable objects into $n_k$ subsets of groups each containing $k$ objects. The associated number of ways is

$$M_3(\vec{n}, A) = \frac{A!}{(1!)^{n_1} n_1! (2!)^{n_2} n_2! (3!)^{n_3} n_3!,..., (A!)^{n_A} n_A!} \tag{2}$$

Using the structure of these two examples as a starting point, a simple generic model can be developed by assigning a weight to each element of a partition $\vec{n} = (n_1, n_2, n_3,..., n_A)$, where $n_k$ is the number of objects of size $k$. A parameter $x_k$ can be introduced into the weight which governs the amount of strength that an element of size $k$ has in the overall weight. Examples of $x_k$ are listed in table 2. For now $x_k$ is left arbitrary. A vector $\vec{x} = (x_1, x_2, x_3,..., x_5)$ is introduced. A weight $W_A(\vec{n}, \vec{x})$ with the structure

$$W_A(\vec{n}, \vec{x}) = \prod_{k=1}^{A} \frac{(x_k)^{n_k}}{n_k!} \tag{3}$$

has some simple features that lead to exactly soluble models with a rich spectrum of results. The $W_A(\vec{n}, \vec{x})$ gives a weight to each partition with a factor $n_k!$ included in the weight. The partition function $Z_A(\vec{x})$ is the weight $W_A(\vec{n}, \vec{x})$ summed over all partitions $\vec{n}$:

$$Z_A(\vec{x}) = \sum_{\vec{n}} W_A(\vec{n}, \vec{x}) = \sum_{\vec{n}} \prod_{k=1}^{A} \frac{(x_k)^{n_k}}{n_k!} \tag{4}$$

To obtain an expectation, such as $<n_j>$, note

$$n_j \prod_{k=1}^{A} \frac{(x_k)^{n_k}}{n_k!} = (x_j) \cdot \frac{(x_j)^{n_j - 1}}{(n_j - 1)!} \prod_{k \neq j}^{A} \frac{(x_k)^{n_k}}{n_k!} \tag{5}$$

Since $n_j! \to (n_j - 1)!$, the operation $n_j W_A(\vec{n}, \vec{x})$ acts to remove one column of $j$ blocks in Fig.1. Thus

$$<n_j> = \frac{\sum_{\vec{n}} n_j W_A(\vec{n},\vec{x})}{Z_A(\vec{x})} = x_j \frac{Z_{A-j}(\vec{x})}{Z_A(\vec{x})} \qquad (6)$$

The $Z_{A-k}(\vec{x})$ arises from summing over all partition with one $j$ column removed in the block picture. Since $A = \Sigma j n_j = \Sigma j <n_j>$ a straightforward rearrangement gives the useful recurrence relation

$$Z_A(\vec{x}) = \frac{1}{A} \sum_{k=1}^{A} k \cdot x_k \cdot Z_{A-k}(\vec{x}) \qquad (7)$$

The $Z_0(\vec{x}) \equiv 1$. The $Z_A(\vec{x})$ is the canonical partition function in statistical mechanics since $A$ is fixed.

The $x_k$ contains the parameters associated with the problem being addressed. A wide range of problems can be addressed using the general model of this section. Some examples are given in Table 2. For example, in statistical physics $x_k$ will depend on thermodynamic variables volume $V$ and temperature $T$. A specific connection of $x_k$ is with the virial expansion of the equation of state which reads

$$\frac{PV}{k_B T} = A - (x_2/x_1^2)A^2 + ((4x_2^2 - 2x_1 x_2)/x_1^4)A^3 + ((-20x_2^3 + 18x_1 x_2 x_3 - 3x_1^2 x_4)/x_1^6)A^4... \qquad (8)$$

The virial expansion also arises from a linked cluster diagrammatic expansion of the partition function [1,7].

The behaviors of $x_k$ with $V$ and $T$ for nearly ideal Bose-Einstein and Fermi-Dirac gases are given in Table 2 along with some selective other examples. The Bose-Einstein case will be discussed further in sec.II.F. The extra $(-1)^{k+1}$ factor in the Fermi-Dirac case arises from antisymmetrization. For fermions a spin degeneracy factor must also be included. The cluster model of Table 2 was originally developed in ref.[35]. The role of a simple binding energy in cluster formation is included through a Boltzmann factor in $x_k$ as listed. The first order phase transition model was introduced in ref.[42,43] to study the liquid-gas phase transition in nuclear systems. The surface energy played an important role in this transition. At the critical point the surface energy coefficient $a_S \to 0$ and the first order transition becomes a second order transition. The stochastic branching model [37] is summarized in Sect.II.C.2. The branching probability is $p$, while the survival probability is $1-p$. The Catalan number counts the number of diagrams with a fixed number of surviving lines. The population genetics model is the neutral allele model of Ewens [44-46] in which genetic drift and mutation are the determining factors in the allelic distribution in a sample. The correspondence between the neutral allele model and a physics model is discussed in ref.[47]. The count distribution examples in Table 2 are discussed in sec.II.B and II.C. The signal to noise model was originally due to Cahill and Glauber [48,49] and is cast in the framework of the results in this section in ref.[36,38].

The signal part describes coherent emission with a Poisson distribution while the noise part arises from chaotic emission process with a negative binomial distribution. The quantum optics signal to noise picture along with several other cases are also discussed in ref[50] as models for particle production in high energy collisions. The microcanonical and canonical descriptions for atoms in a laser trap or harmonic oscillator trap were developed in ref.[51].

Table 2  Examples of $x_k$

| Area | $x_k$ | |
|---|---|---|
| Bose-Einstein gas | $\dfrac{L^d}{\lambda_T^d}\dfrac{1}{k \cdot k^{d/2}}$, | dimension $d$, $L$ length, $\lambda_T = h/\sqrt{2\pi m k_B T}$ |
| Fermi-Dirac gas | $(-1)^{k+1}\dfrac{L^d}{\lambda_T^d}\dfrac{1}{k \cdot k^{d/2}}$ | |
| Cluster Model | $\dfrac{V}{\lambda_T^3}\dfrac{1}{k}\exp(-\dfrac{a_V}{k_B T})$, | binding energy of a cluster : $E_b = a_V(k-1)$ |
| $1^{st}$ order phase transition | $(V/\lambda_T^3)k^{3/2} \cdot$ $\exp(\beta(a_v k - a_S k^{2/3}))$ | $\beta = 1/k_B T$, $a_S$ surface energy coefficient $E_b = a_V k - a_S k^{2/3}$ |
| Stochastic model of branching | $p^{k-1}(1-p)^k \widehat{C}[k]$ | $p$ branching & $1-p$ survival probabilities $\widehat{C}[k]$ Catalan number: |
| Population genetics allelic distributions | $4N_e \mu / k$ | $N_e$ effective population size $\mu$ mutation rate |
| Particle count distributions | $xt^k / k$ | negative binomial $<A> = xt/(1-t)$ $<A^2> - <A>^2 = <A> + <A>^2/x$ |
| Glauber's photon count model | $1/2^{2(k-1)} x\widehat{C}[k]xt^k$ $\to xt^k/k^{3/2}$ | $<A> = xt/\sqrt{1-t}$ $<A^2> - <A>^2 = <A> + <A>^2/2x\sqrt{1-t}$ |
| Signal (S) to Noise (Nl) | $(y + \dfrac{x}{k})t^k$ | $y = \dfrac{aS}{Nl(1+Nl/a)}$, $t = \dfrac{Nl/a}{1+Nl/a}$ |
| Atoms in a laser trap | $\dfrac{1}{k}\prod_i \dfrac{\sqrt{\Omega_i}}{1-\Omega_i^k}$ | $\Omega_i = \exp(-\beta\hbar\omega_i)$, $i = x, y, z$ |

Polya theory is a standard tool of enumerative combinatorics. The approach uses generating functions based primarily on permutations and its associated cycle class decomposition. A connection of Polya urn models with the theory of disordered systems was also developed in ref.[52]. A generating function for the above generic model will be considered next. The generating function

$$Z_{gf}(\vec{x}, u) = Exp[x_1 u + x_2 u^2 + x_3 u^3 + x_4 u^4 + x_5 u^5 + .....] \tag{9}$$

when expanded as a series involving powers of $u$ with terms with the same power of $u$ collected together gives the following:

$$Z_{gf}(\vec{x}, u) = \sum_{n=0}^{\infty} \frac{(x_1 u + x_2 u^2 + x_3 u^3 + x_4 u^4 + x_5 u^5 + .....)^n}{n!} =$$

$$1 + x_1 u + \{\frac{x_1^2}{2!} + x_2\} u^2 + \{\frac{x_1^3}{3!} + \frac{2 x_1 x_3}{2!} + x_3\} u^3 + \tag{10}$$

Thus

$$Z_{gf}(\vec{x}, u) = \sum_{A=0}^{\infty} Z_A(\vec{x}) u^A \tag{11}$$

The $Z_{gf}(\vec{x}, u)$ is the exponential generating function for the canonical partition function $Z_A(\vec{x})$. The $Z_{gf}(\vec{x}, u)$ is referred to as the grand canonical partition function in statistical mechanics with $u$ the fugacity. The fugacity $u = \exp[\beta \mu]$ with $\mu$ the chemical potential and $\beta = 1/k_B T$, where $T$ is the temperature in Kelvin and $k_B$ is the Boltzmann constant.

Another important quantity is the probability that $k$ is the largest object in ensemble [53]. We associate the symbol $k_L$ when discussing this $k$. This probability can be obtained from the partition functions. Using the symbol $P_A(k_L)$ for this probability, the $P_A(k_L)$ can be obtained from $Z_A(\vec{x})$

$$P_A(k_L) = \frac{Z_A(x_1,..., x_{k_L}, \{x_{k_L+1} = 0,...x_A = 0\}) - Z_A(x_1,..., x_{k_L-1}, \{x_{k_L} = 0,..., x_A = 0\})}{Z_A(x_1, x_2,...., x_A)} \tag{12}$$

In this expression, all the $x_j's$ in the { } bracket are set equal to zero. The first term in the numerator guarantees that no higher $k$ than $k_L$ is present in the partition. The second term arises from the fact that $k_L$ itself might not be present in the partition even though no higher $k$ is present. Therefore, those partitions must be also subtracted out to give a result that has $k_L$ present. The $Z_A(x_1,...,x_{k_L},\{x_{k_L+1}=0,...x_A=0\})$ can be calculated from the recurrence relation for generating the partitions $Z_A(\vec{x})$ by using the set of $x_k's$ given by: $(x_1,..,x_{k_L},\{x_{k_L+1}=0,..,x_A=0\})$. A similar procedure holds for $Z_A(x_1,..,x_{k_L-1},\{x_{k_L}=0,...,x_A=0\})$ with $(x_1,..,x_{k_L-1},\{x_{k_L}=0,...,x_A=0\})$. Setting $\{x_{k_L+1}=0,...,x_A=0\}$ means we have also removed all $n_k$ with $k \geq k_L$ from $\vec{n}$. The $\vec{n}$ will then span all partitions with no part greater than $k_L$. $Z_A(x_1,...,x_{k_L},\{x_{k_L+1}=0,...x_A=0\})$ is the partition function associated with this reduced or limited $\vec{n}$. The expectation that $k_L$ is the largest object is then:

$$<k_L> = \sum_{k_L=1}^{A} k_L P_A(k_L) \qquad (13)$$

The mean square fluctuation can also be calculated from $P_A(k_L)$. The behavior of the largest cluster is useful when discussing phase transitions [42,43] where a large cluster suddenly appears. The sudden appearance of a large cluster parallels the sudden appearance of the "infinite" cluster in percolation models.
If $u$ is set equal to 1, then

$$Z_{gf}(\vec{x},u=1) = Z_{gf}(\vec{x},1) = \sum_{A=0}^{\infty} Z_A(\vec{x}) \qquad (14)$$

A probability distribution can also be developed from $Z_A(\vec{x})$ & $Z_{gf}(\vec{x},1)$ as

$$P_A(\vec{x}) = \frac{Z_A(\vec{x})}{Z_{gf}(\vec{x},1)} \qquad (15)$$

The $<A>$ can be obtained from

$$<A> = \frac{\partial Z_{gf}(\vec{x},u)/\partial u}{Z_{gf}(\vec{x},u)}\bigg|_{u \to 1} = \sum_{k=1}^{\infty} k x_k = \sum_{A=0}^{\infty} A P_A(\vec{x}). \qquad (16)$$

Similarly, the fluctuation is the second moment of the $x_k's$ given by

$$<A^2> - <A>^2 = \sum_{k=1}^{\infty} k^2 x_k \qquad (17)$$

The probability for $A=0$, referred to as the void probability, is $P_0(\vec{x}) = 1/Z_{gf}(\vec{x},1)$.
The $P_A(\vec{x})$ can represent photon count probabilities from a laser as in the Glauber model listed in Table 2 or particle count probabilities from a high energy collision.

II.B. Permutation Model:

$$x_k = \frac{1}{k} x t^k \tag{18}$$

For certain choice of $x_k$, the $Z_A(\vec{x})$ takes on a simple form. One such case, rooted in the theory of permutations, involves $M_2$ given above. Results obtained from this choice also lead to negative binomial distribution, which is of interest because it has large non-poissonian fluctuations. The $M_2$ occurs as part of the weight in Feynman's approach to Bose-Einstein condensation [1] which is also discussed in Sect.II.F. Expressions developed in this section also have their counterparts in population genetics where $n_k$ is the number of different genes each occurring $k$ times in a sample of size $A$ as in the sampling theory of Ewens. The Ewens sampling weight was also used extensively in a model of aggregate behavior in economics by Aoki [54]. An exactly soluble model for clustering or fragmentation processes can be developed using the special case of $x_k$ [35] listed in Table 2.

The permutation based model has $x_k = xt^k/k$ which leads to a weight structure

$$W_A(\vec{n}, \vec{x}) = \prod_{k=1}^{A} \frac{x^{n_k} t^{k \cdot n_k}}{n_k! k^{n_k}} = t^A \prod_{k=1}^{A} \frac{x^m}{n_k! k^{n_k}} \tag{19}$$

The $t^A$ is common to all partition. The multiplicity $m = \Sigma n_k$ and gives the same weight to all partitions with the same $m$. For $x > 1$, partitions with large $m$ are favored and for $x < 1$ partitions with small $m$ are favored. The associated $Z_A(\vec{x})$ is

$$Z_A(\vec{x}) = t^A \frac{x(x+1)....(x+A-1)}{A!} = \frac{\Gamma(x+A)}{\Gamma(x)\Gamma(A+1)} t^A = (-1)^{A-1} U[-A, x, 0] \frac{t^A}{A!} \tag{20}$$

The $\Gamma(x)$ is a gamma function with the recurrence property $\Gamma(x+1) = x\Gamma(x)$ and $\Gamma(A+1) = A!$. The $U[-A, x, 0]$ is a confluent hypergeometric function. A closely related quantity is the Pochhammer symbol $[x]_A = x(x+1)(x+2).....(x+A-1) = \Gamma(x+A)/\Gamma(x)$
A single term recurrence relation holds for this model $Z_A(\vec{x}) = (x+A-1)Z_{A-1}(\vec{x})t/A$.

The generating function (with $t = 1$) is

$$Z_{gf}(\vec{x}, u) = Exp[xu + \frac{x}{2}u^2 + \frac{x}{3}u^3 + .....] = Exp[-x \cdot \ln(1-u)] = \frac{1}{(1-u)^x} =$$
$$Exp[xu(_2F_1[1,1,2,u])] \tag{21}$$

using $\ln(1-u) = -u - u^2/2 - u^3/3...$ . The $_2F_1[1,1,2,u] = -\ln[1-u]/u$ is a Gauss hypergeometric function with $a = 1, b = 1, c = 2$. The general Gauss hypergeometric function has an expansion

$$_2F_1[a,b,c,z] = \sum_{n=0}^{\infty} \frac{[a]_n [b]_n}{[c]_n n!} z^n \tag{22}$$

The Gauss hypergeometric function can form the basis for a generating function for the canonical partition functions $Z_A$ associated with it as discussed in II.F. The $<n_k>$ is

$$<n_k> = \frac{x}{k} \frac{A!}{(A-k)!} \frac{x(x+1).....(x+(A-k)-1)}{x(x+1)....(x+A-1)} \tag{23}$$

At $x=1$

$$<n_k> = \frac{1}{k} \tag{24}$$

which is a pure power law fall off only restricted by the constraint $k \leq A$. In general, a power law distribution can fall with size $k$ with a power $\tau$ so that

$$<n_k> \sim \frac{1}{k^\tau} \tag{25}$$

The probability distribution associated with the permutation model is

$$P_A(\vec{x}) = \frac{Z_A(\vec{x})}{Z_{gf}(\vec{x},1)} = \frac{\frac{\Gamma(x+A)}{\Gamma(x)\cdot\Gamma(A+1)} t^A}{\frac{1}{(1-t)^x}} = \binom{x+A-1}{A} t^A (1-t)^x \tag{26}$$

The mean $<A> = xt/(1-t)$ and variance $<A^2> - <A>^2 = <A>(1+<A>/x)$.
The variance is that of a negative binomial (NB) distribution. Solving for $t$ in terms of $<A>/x$ and substituting into $P_A(\vec{x})$ gives another form for $P_A(\vec{x})$:

$$P_A(\vec{x}) = \binom{x+A-1}{A} \left(\frac{<A>/x}{1+<A>/x}\right)^A \frac{1}{(1+<A>/x)^x} = \binom{x+A-1}{A} (1-p)^A p^x \tag{27}$$

The limit $x \to 1$ results in the Planck distribution.
 The NB distribution has been an important distribution in particle physics were it has been used to discuss intermittency and fractal structure [55,56,57] and shown to arise from a self-similar cascading process.

II.C.1 Catalan Model:
$$x_k = \frac{1}{2^{2(k-1)}} \frac{1}{k} \frac{(2k-2)!}{(k-1)!(k-1)!} xt^k = \frac{1}{2^{2(k-1)}} \hat{C}[k] xt^k \tag{28}$$

 A second case will now be given which has connections with several areas of physics such as in quantum optics and particle physics. In quantum optics, results from this second case appear in photon count distributions operating near a laser threshold. This photon count model was developed by Glauber [39,40]. As already noted results from this case can be connected to a Feynman-Wilson gas used to understand rapidity distributions in particle collisions. Photon and particle count distributions follow from the probability interpretation given by $P_A(\vec{x}) = Z_A(\vec{x})/Z_{gf}(\vec{x},1)$. The negative binomial (NB) probability was already discussed in the previous subsection. The $x_k$ has a component that involves the Catalan numbers in mathematics. The Catalan numbers are consider by some mathematician as the second most popular set of numbers with the first being the Fibonacci numbers. The $\widehat{C}(k) \equiv (1/k)\cdot(2k-2)!/((k-1)!)^2$ is a shifted Catalan number: $\widehat{C}(1)=1, \widehat{C}(2)=1, \widehat{C}(3)=2, \widehat{C}(4)=5, \widehat{C}(5)=14, \widehat{C}(6)=42,\ldots$ .For large $k$, the asymptotic behavior of $\widehat{C}[k] \to 2^{2(k-1)}/\sqrt{\pi}k^{3/2}$. The Catalan numbers are $C = 1,2,5,14,42,\ldots$ for $n = 1,2,3,4,5,\ldots$. For this choice of $x_k$ the partition functions are

$$Z_{gf}(\vec{x},u) = Exp[2x(1-\sqrt{1-tu})] = Exp[xut(_2F_1[1/2,1,2,tu])] \tag{29}$$

The $_2F_1[1/2,1,2,z] = 2(1-\sqrt{1-z})/z$ is a hypergeometric function with $a = 1/2, b = 1, c = 1$. The associated canonical partition function is $Z_A$ and it is given by

$$Z_A(\vec{x}) = (2x)^{2A}U[A,2A,4x]\frac{t^A}{A!} = (2x)^{2A}\frac{t^A}{A!}\left(\frac{4^{\frac{A}{2}}}{\sqrt{\pi}}\exp(2x)x^{\frac{1}{2}-A}K_{A-1/2}(2x)\right) \tag{30}$$

The $U[A,2A,2x]$ is a confluent hypergeometric function, while $K_{A-1/2}(2x)$ is a Bessel $K$ function of fractional order $A-1/2$. The Glauber photon count model is just the ratio $P_A = Z_A(\vec{x})/Z_{gf}(\vec{x},1)$.

Defining $Q_A = A! \cdot Z_A$ the $Q_A$ satisfy a simple two term recurrence that reads

$$Q_{A+1} = \frac{1}{2}(2A-1)Q_A + x^2 Q_{A-1} \tag{31}$$

II.C.2 Branching Model:

$$x_k = \frac{1}{2^{2(k-1)}}\frac{1}{k}\frac{(2k-2)!}{(k-1)!(k-1)!}xt^k \to \beta p^{k-1}(1-p)^k \widehat{C}[k] \tag{32}$$

$$x = \beta/4p, \quad t = 4p(1-p) \tag{33}$$

A branching model [37] can be developed around the Catalan model by introducing a branching probability or stochastic ancestral variables which are $x = \beta/4p$, $t = 4p(1-p)$. The branching or lines of descent interpretations of $p^{k-1}(1-p)^k\widehat{C}[k]$ are shown in Fig.2

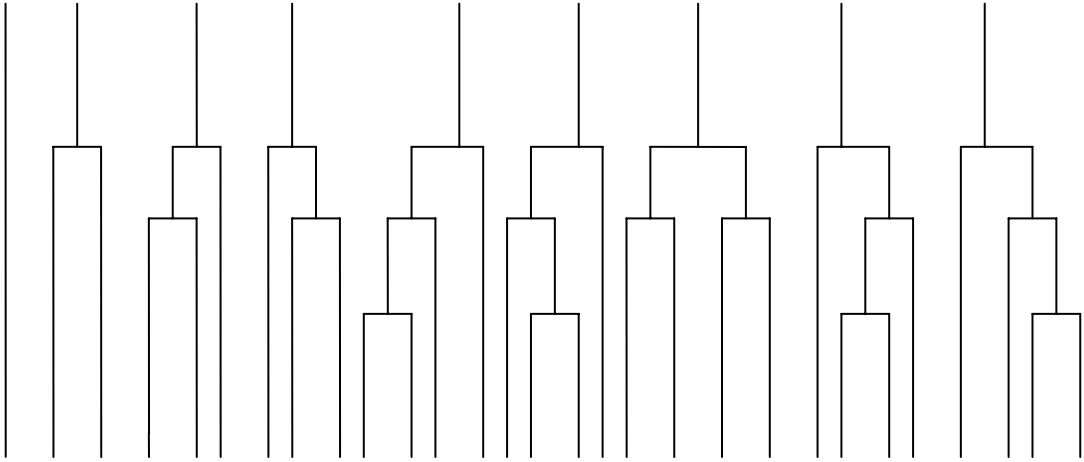

Fig.2 Branching or evolutionary lines of descent in an hierarchical topology. The Catalan number $\hat{C}[k]$ counts the number of diagrams arising from the same number of bifurcations. Bifurcations occur with probability $p$ while the probability of survival of a line is $(1-p)$.

The sum $\Sigma_k x_k = \beta$ for all $p \le 1/2$. For $p \ge 1/2$, $\Sigma_k x_k = \beta(1-p)/p$. The value $p = 1/2$ acts as a critical transition point.

 II.D Levy Index Model:

$$x_k = \frac{[a]_{k-1}}{k!} xt^k \tag{34}$$

The permutation and Catalan models are special cases of

$$x_k = \frac{[a]_{k-1}}{k!} xt^k = \frac{\Gamma(a+k-1)}{\Gamma(a)\Gamma(k+1)} xt^k \tag{35}$$

with $a = 1$ for the permutation model and $a = 1/2$ for the Catalan model. For large $k$

$$x_k \sim \frac{1}{\Gamma(a)} \frac{1}{k^{2-a}} xt^k \tag{36}$$

The grand canonical partition function or generating function is

$$Z_{gf}(\vec{x},u;a) = Exp[xu({}_2F_1(a,1;2,u))] = Exp[\frac{x}{a-1}(\frac{1}{(1-u)^{a-1}}-1)] \tag{37}$$

The associate canonical partition functions are easily developed from the recurrence relation

$$Z_A(\vec{x}) = \frac{1}{A}\sum_{k=1}^{A} k \cdot \frac{[a]_{k-1}}{k!} x \cdot Z_{A-k}(\vec{x}) \qquad (38)$$

The overall $t^A$ factor in each $Z_A$ is omitted. Further generalized models based on $xu(_2F_1(a,b;c,u))$ are discussed in subsection II.F.

II.E Fibonacci/Lucas Models
   II.E.1 Case1:

$$x_k = (\frac{2}{\sqrt{5}+1})^k \frac{1}{k} L_k xt^k = (\phi')^k \frac{1}{k} L_k xt^k \qquad (39)$$

The Fibonacci numbers are generated by either a recurrence relation or a series expansion. Let $F_n$ be the $n$'th Fibonacci number. The recurrence expansion is $F_n = F_{n-1} + F_{n-2}$ with $F_0 = F_1 = 1$, the series reads $1,1,2,3,5,8,13,21,\ldots$.
The Fibonacci series is also seen as coefficients in the expansion of a generating function given by

$$\frac{1}{1-z-z^2} = 1 + z + 2z^2 + 3z^3 + 5z^4 + 8z^5 + 13z^6 + 21z^7 + \ldots \qquad (40)$$

The ratio of two adjacent Fibonacci number tends to the Golden Mean as $n$ becomes large: $F_{n+1}/F_n \to (\sqrt{5}+1)/2 = 1.61803$. A generalized Fibonacci series is defined as $G_{n+1} = G_n + G_{n-1}$ with any two starting points $G_0, G_1$. Lucas numbers $L_n$ have the starting point $G_0 = 2, G_1 = 1$ to generate the series: $2,1,3,4,7,11,18,\ldots$. If the natural logarithm of the generating function is taken then an interesting feature appears. Namely:

$$Ln[1/(1-u-u^2)] = \frac{1}{1}u + \frac{3}{2}u^2 + \frac{4}{3}u^3 + \frac{7}{4}u^4 + \frac{11}{5}u^5 + \frac{18}{6}u^6 + \ldots = \sum_{n=1}^{\infty}\frac{L_n}{n}u^n \qquad (41)$$

where the coefficients satisfy a Fibonacci recurrence relation $L_n = L_{n-1} + L_{n-2}$ with initial values $L_1 = 1, L_2 = 3$. The starting points of any generalized Fibonacci series do not affect the asymptotic ratio $G_{n+1}/G_n \to (\sqrt{5}+1)/2$. The $L_0 = 2$ point will be excluded and instead $L_1 = 1, L_2 = 3$ will be used as the starting point of the Lucas series. The $L_n$ have the property $L_n/((\sqrt{5}+1)/2)^n \to 1$. For the Fibonacci series $F_n$ with starting points $F_0 = 1, F_1 = 1$ the ratio $F_n/((\sqrt{5}+1)/2)^n \to (\sqrt{5}+1)/2\sqrt{5}) = .7726$. The $F_n$ is given by $F_n = ((1+\sqrt{5})/2\sqrt{5})r_+^n - ((1-\sqrt{5})/2\sqrt{5})r_-^n$. The $r_\pm = (1\pm\sqrt{5})/2$ are the two solutions of the quadratic equation $r^2 - r - 1 = 0$ which arise from the recurrence relation $F_n = F_{n-1} + F_{n-2}$ assuming $F_n/F_{n+1} \to r$. Similarly, $L_n = r_+^n + r_-^n$.

A choice for $x_k$ that reads

$$x_k = (\frac{2}{\sqrt{5}+1})^k \frac{1}{k} L_k xt^k = (\phi')^k \frac{1}{k} L_k xt^k \tag{42}$$

has features that parallel the permutation case $x_k = xt^k/k$, both falling as $1/k$ and each generating a log series in $\Sigma x_k u^k$. The $Z_{gf}(\vec{x},u)$ is

$$Z_{gf}(\vec{x},u) = \exp \sum_{k=1}^{\infty}(\phi')^k \frac{1}{k} L_k xt^k u^k = \exp[x \ln[1/(1-\phi'tu-(\phi't)^2 u^2)]]$$

$$= (1-(\phi't)u-(\phi't)^2 u^2)^{-x} = \sum_{A=0}^{\infty} Z_A(\vec{x}) u^A \tag{43}$$

Note also the result (setting $\phi't = 1$ for now)

$$\frac{1}{(1-z-z^2)^x} = \frac{1}{(1-r_+z)^x} \frac{1}{(1-r_-z)^x} \tag{44}$$

The individual factors $1/(1-r_+z)^x$ & $1/(1-r_-z)^x$ are each generating functions for the permutation model with $u \to r_+z$ & $u \to r_-z$. Consequently

$$Z_{gf}(\vec{x},u) = \exp[x \ln[1/(1-\phi'tu-(\phi't)^2 u^2)]] = \exp\{x \ln[(1/(1-r_+\phi'tu) + x \ln[1/(1-r_-\phi'tu)]\}$$

$$= \exp\{xtu(_2F_1[1,1,2,tu] - (\phi')^2 \cdot _2F_1[1,1,2,-(\phi')^2 tu])\} \tag{45}$$

where we have used $r_+\phi' = 1$ and $r_-\phi' = -(\phi')^2$. Thus, the generating function or grand canonical potential can once again be written in terms of Gauss hypergeometric functions. For this case two hypergeometric function are necessary with both having $a=1, b=1, c=2$. The Fibonacci/Lucas canonical ensemble can be written as a convolution of two permutation canonical ensembles. Writing

$$\frac{1}{(1-r_\pm u)^x} = \sum_{n=0}^{\infty}(Z_n^\pm(x,r_\pm))u^n \tag{46}$$

the

$$Z_n^+(x,r_+) = \frac{x(x+1)(x+2)....(x+n-1)}{n!} r_+^n = \frac{\Gamma(x+n)}{\Gamma(x)\Gamma(n+1)} r_+^n \tag{47}$$

and its dual is

$$Z_n^-(x,r_-) = \frac{x(x+1)(x+2)....(x+n-1)}{n!}(r_-)^n = \frac{\Gamma(x+n)}{\Gamma(x)\Gamma(n+1)} r_-^n \qquad (48)$$

Thus

$$\frac{1}{(1-u-u^2)^x} = \sum_{A=0}^{\infty} Z_A(x)u^A = (\sum_{n=0}^{\infty} Z_n^+(x,r_+)u^n) \cdot (\sum_{n=0}^{\infty} Z_n^-(x,r_-)u^n)$$
$$= \sum_{A=0}^{\infty} \left( \sum_{m=0}^{A} Z_m^+(x,r_+) \cdot Z_{A-m}^-(x,r_-) \right) u^A \qquad (49)$$

Listed below are a few cases starting with $Z_0(x) = 1$

$$Z_1(x) = \phi'tx$$

$$Z_2(x) = (\phi't)^2 x(x^1 + 3 \cdot 1!)/2!$$

$$Z_3(x) = (\phi't)^3 x(x^2 + 9 \cdot x^1 + 4 \cdot 2!)/3! \qquad (50)$$

$$Z_4(x) = (\phi't)^4 x(x^3 + 18 \cdot x^2 + 59 \cdot x^1 + 7 \cdot 3!)/4!$$

Since the $\phi't$ factor is easy to include at the end, a $\hat{Z}_A(x) \equiv Z_A(x)/(\phi't)^A$ is introduced. A two term recurrence relation exists for $\hat{Z}_A(x)$:

$$\hat{Z}_A(x) = \frac{1}{A}\{(x+A-1)\hat{Z}_{A-1}(x) + (2x+A-2)\hat{Z}_{A-2}(x)\} \qquad (51)$$

Defining $Q_A(x) \equiv A!Z_A(x)/(\phi't)^A$ this last recurrence relation becomes

$$Q_A(x) = \{(x+A-1)Q_{A-1}(x) + (A-1)(2x+A-2)Q_{A-2}(x)\} \qquad (52)$$

The $\hat{Z}_A$ is simply related to the Gegenbauer polynomial $C_A^{(x)}(y)$ which will be shown in the next subsection using an associated generating function. The connection is

$$\hat{Z}_A = i^A C_A^{(x)}(1/2i) \qquad (53)$$

The $i = \sqrt{-1}$. At $x = 1$, the $\hat{Z}_A$ is just the Fibonacci number $F_A$ so that

$$\frac{Z_A(x=1)}{(\phi't)^A} = F_A \qquad (54)$$

For $Z_0(x) = 1, F_0 = 1$. The connection between $Z_A(x=1)$ and $F_A$ can easily be seen from the generating function $Z_{gf}(\vec{x}, u)$.

At $x = 1$, the

$$<n_k> = \frac{1}{k}\frac{L_k F_{A-k}}{F_A} \tag{55}$$

If $A >> 1, A - k >> 1$ then $F_{A-k}/F_A \to 1/\phi^k$ and if $k >> 1$, then $L_k \to \phi^k$. Consequently, under these conditions

$$<n_k> = \frac{1}{k}\frac{L_k F_{A-k}}{F_A} \to \frac{1}{k} \tag{56}$$

This equation is one of the important results of this paper. Specifically, a pure scale invariant power law is encountered with an exponent $\tau = 1$ over the region of $k$ stated. Note, some departure from a strict $1/k$ behavior occurs for small $k \sim 1$ and large $k \sim A$. The behavior of $<n_k>$ at $x = 1$ is shown in Fig. 3 which is a plot of $k<n_k>$ versus $k$.

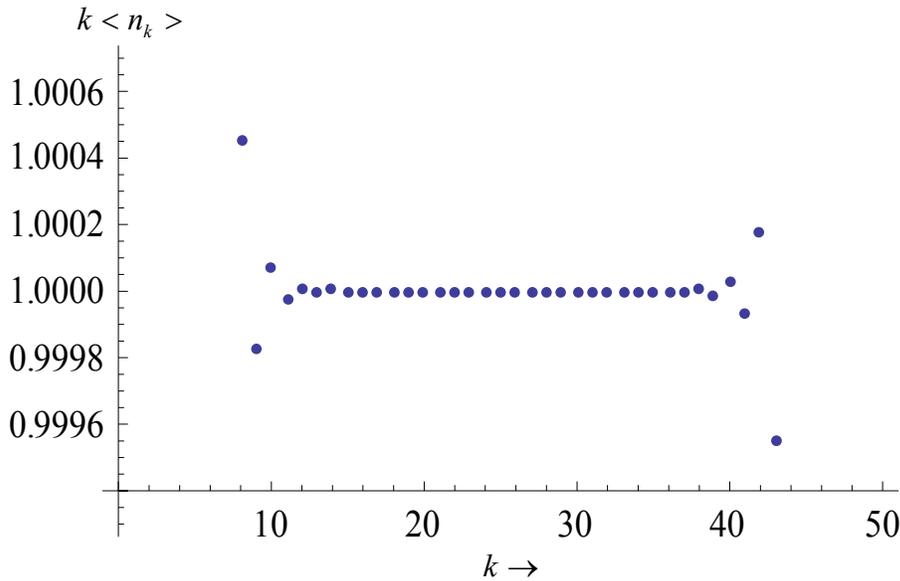

Fig.3 The near hyperbolic behavior of $k<n_k>$. The size of $A = 50$. Departures from $<n_k> = 1/k$ occur at the endpoints. For $k \sim 1-10$, the departure from unity arises from the Lucas numbers not scaling as $\phi^k$, and for $k \sim 40-50$, the departures arise from the ratio of Fibonacci numbers $F_{A-k}/F_A$ not scaling as $\phi^{-k}$.

The sum rule $\Sigma k <n_k> = A$ leads to a convolutions between the Lucas series $L_n$ and the Fibonacci series $F_n$ that reads

$$A = \sum_{k=1}^{A} \frac{L_k F_{A-k}}{F_A} = \frac{L_1 F_{A-1} + L_2 F_{A-2} + L_3 F_{A-3} + ... + L_{A-1} F_1 + L_A F_0}{F_A} \tag{57}$$

with $F_0 = 1, F_1 = 1$ and with $L_1 = 1, L_2 = 3$.

Behavior away from the critical point $x = 1$ is also easily evaluated. If $x = 2$ and setting $\phi't = 1$ again for now, then the generating function $Z_{gf}$ is

$$Z_{gf}(\vec{x}, u) = \frac{1}{(1 - u - u^2)^2} = (F_0 + F_1 u + F_2 u^2 + F_3 u^3 + .....)^2 =$$

$$F_0 F_0 + (F_0 F_1 + F_1 F_0) u + (F_0 F_2 + F_1 F_1 + F_2 F_0) u^2 + ... \tag{58}$$

and therefore

$$Z_A(x = 2) = (\phi't)^A \sum_{j=0}^{A} F_j F_{A-j} = (\phi't)^A \, i^A C_A^{(2)}(1/2i) \tag{59}$$

Then

$$< n_k > = x_k \frac{Z_{A-k}}{Z_A} = \frac{2}{k} L_k \frac{\sum_{j=0}^{A-k} F_j F_{A-k-j}}{\sum_{j=0}^{A} F_j F_{A-j}} \tag{60}$$

The constraint sum rule $A = \Sigma k < n_k >$ leads to

$$A \sum_{j=0}^{A} F_j F_{A-j} = 2 \sum_{k=1}^{A} L_k \sum_{j=0}^{A-k} F_j F_{A-k-j} \tag{61}$$

Also, the convolution

$$\sum_{j=0}^{A} F_j F_{A-j} = \frac{1}{5}(A+1) L_{A+2} + \frac{2}{5} F_A \tag{62}$$

which follows from the $r_+, r_-$ expressions for $F_j$. Thus at $x = 2$

$$< n_k > = x_k \frac{Z_{A-k}}{Z_A} = \frac{2}{k} L_k \frac{(A-k) L_{A-k+2} + 2 F_{A-k}}{A L_{A+2} + 2 F_A} \tag{63}$$

Fig.4 illustrates the behavior of $<n_k>$ at $x=2$. At this value of $x$ and for all $x>1$, the $<n_k>$ decrease faster with $k$ than the power law $1/k$. For $x>1$, high multiplicity partitions are favored. Fig.4 shows nearly a straight line behavior except for the initial points. The result follows from

$$k<n_k> = \frac{2(A-k)L_{A-k+2}L_k + 2F_{A-k}L_k}{AL_{A+2} + 2F_A} \tag{64}$$

and properties of Lucas and Fibonacci numbers. Namely, the result $L_{A-k+2}L_k/L_A \cong 1$, except for $k \sim 1, k \sim A$, determines the straight line behavior. The $x=3,4,\ldots$ no longer appear as straight lines. However, the extrapolated $k<n_k>$ intercept is the value of $x$ for small $k$. Other integer choices of $x$, $x=3,4,\ldots$ lead to higher order relations.

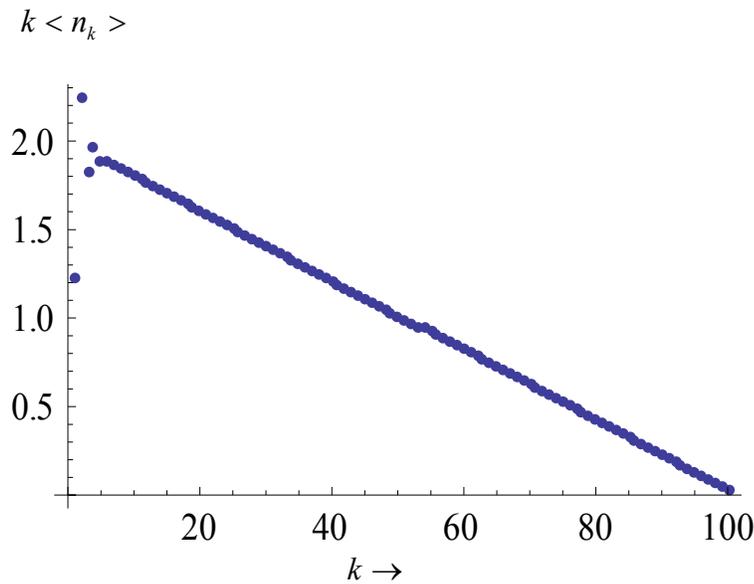

$k<n_k>$

$k \rightarrow$

Fig.4 . Plot of $k<n_k>$. A value $A=100$ was used and $x=2$. The behavior of $k<n_k>$ is a basically a straight line with slope -2/100 for this $x$. A exact straight line behavior leads to $k<n_k> = (2(A-k)/A)+1/A$. The $1/A$ term is added so the constraint $\Sigma k<n_k> = A$ is satisfied.

The Gegenbauer polynomial $C_A^{(x)}(y)$ at $x=1/2$ is the Legendre polynomial $P_A(y)$: $C_A^{(1/2)}(y) = P_A(y)$. Fig.5 shows the $U$ shaped behavior of $<n_k>$. The $U$ arises from the fact that a large cluster has associated with it small clusters because of the constraint. The mass fraction $k<n_k>/A$ or $k<n_k>$ is in large clusters as shown in Fig.6

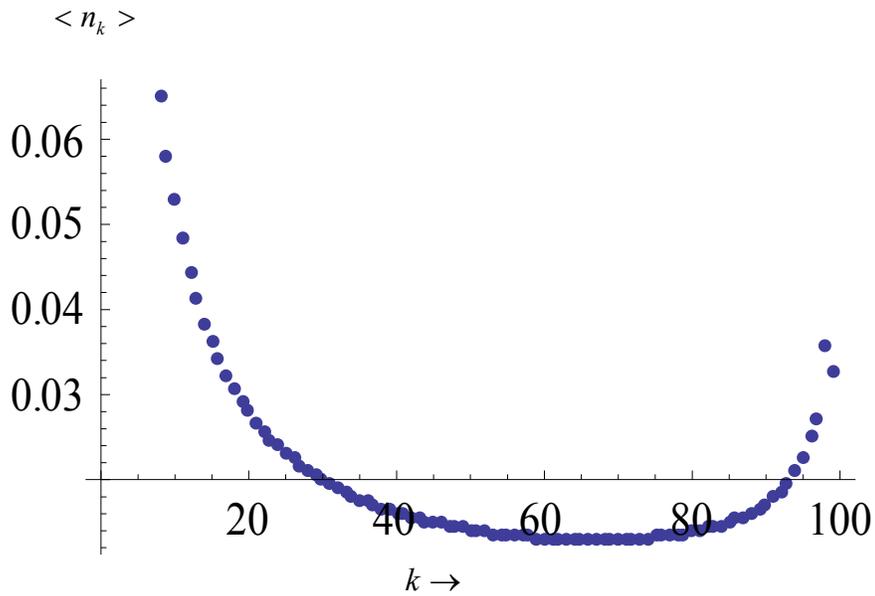

Fig.5. Plot of $<n_k>$ for $x = 1/2$ and $A = 100$. For $x < 1$ large clusters appear and the distribution $<n_k>$ has a $U$ shape. Large clusters are accompanied with small clusters. A cluster with size $k = A-1$ is accompanied by a $k = 1$ monomer; a cluster with size $k = A-2$ has either two monmers or one dimer.

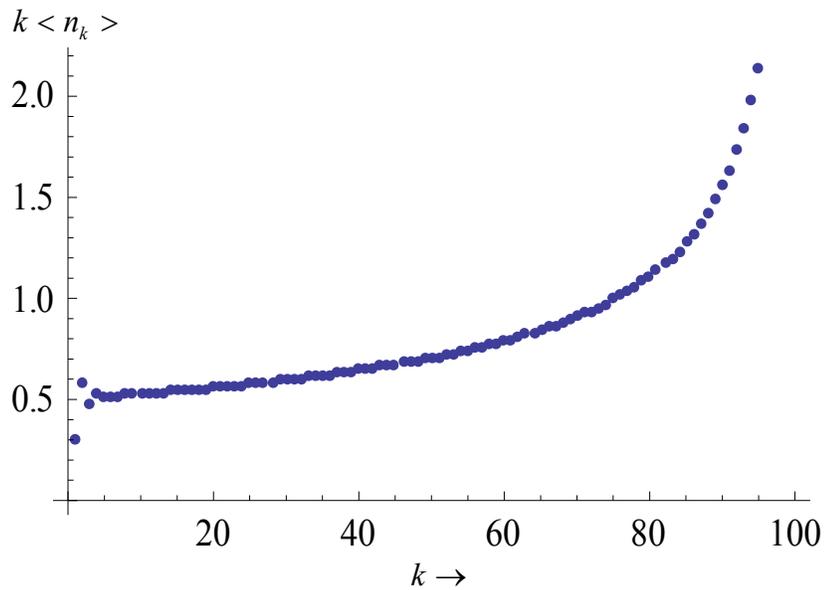

Fig.6. Plot of $k<n_k>$ versus $k$ for $x = 1/2$ and $A = 100$. The mass fraction $k<n_k>/A$ is in very large clusters. For $x < 1$, low multiplicity partitions are favored.

II.E.2 Case 2, Fibonacci/Lucas Polynomials:

$$x_k = \frac{1}{k}(g_+^k(y) + g_-^k(y)) = x\frac{1}{k}L_k(y) \quad g_\pm(y) = \frac{y \pm \sqrt{y^2+4}}{2} \tag{65}$$

A set of polynomials $F_n(y)$, called the Fibonacci polynomials, are defined as the coefficient of $u^n$ in the expansion of $1/(1-yu-u^2)$ giving

$$1/(1-yu-u^2) = \sum_{n=0}^{\infty} F_n(y)u^n =$$
$$1 + yu + (1+y^2)u^2 + (2y+y^3)u^3 + (1+3y^2+y^4)u^4 + (3y+4y^3+y^5)u^5 + \ldots \tag{66}$$

The Fibonacci polynomials satisfy the recurrence relation $F_n(y) = yF_{n-1}(y) + F_{n-2}(y)$. At $y=1$ the coefficients of $u$ are the Fibonacci numbers $F_n$ with the 1,1 starting point. The $F_n(y)$ are also simply related to the Chebyshev polynomial $U_n(y)$ of the second kind which is a special case of the Gegenbauer polynomial $C_n^{(1)}(y/2i)$. Specifically

$$F_n(y) = (i^n)U_n(y/2i) = (i^n)C_n^{(1)}(y/2i) \tag{67}$$

The Fibonacci polynomials are also connected to binomial coefficients as

$$F_A(y) = \sum_{j=0}^{[A/2]} \binom{A-j}{j} y^{A-2j} \tag{68}$$

where $[A/2] = (A-1)/2$ for $A$ odd. The $[A/2]$ is the largest integer in $A/2$. The generating function $1/(1-yu-u^2)$ can be developed from an exponential generating function. Using $1-yu-u^2 = (1-g_+(y)u)\cdot(1-g_-(y)u)$ the

$$\frac{1}{1-yu-u^2} = \exp[\{\ln 1/(1-g_+(y)u) + \ln 1/(1-g_-(y)u)\}]$$
$$= \exp[\{ug_+(y)\cdot_2F_1[1,1,2,g_+(y)u] + ug_-(y)\cdot_2F_1[1,1,2,g_-(y)u]\}]$$
$$= \exp[\sum_{k=1}^{\infty} \frac{(g_+(y))^k}{k}u^k + \sum_{k=1}^{\infty} \frac{(g_-(y))^k}{k}u^k] = \exp[\sum_{k=1}^{\infty} x_k u^k] \tag{69}$$

giving $x_k = L_k(y)/k$. At $y=1$, $g_\pm = (1 \pm \sqrt{5})/2 = r_\pm$.
The $L_k(y)$ is the Lucas polynomial. The $L_k(y)$ is related to Fibonacci polynomials through the recurrence $L_k(y) = yF_{k-1}(y) + 2F_{k-2}(y)$ or $L_k(y) = F_k(y) + F_{k-2}(y)$. The $F_0(y) = 1, F_1(y) = y, F_2(y) = y^2+1,\ldots$, and $L_1(y) = y$, $L_2(y) = y^2+2$, $L_3(y) = y^3+3y$. The ratio of $F_A(y)/F_{A-1}(y) \to g_+(y)$ as $A \to \infty$ with the same result for the Lucas polynomials

The above result for the generating function result can be easily extended to read

$$\frac{1}{(1-yu-u^2)^x} = \exp[\sum_{k=1}^{\infty} x \frac{1}{k}(g_+^k(y) + g_-^k(y))u^k] = \sum_{A=0}^{\infty} Z_A(x,y)u^A \qquad (70)$$

For arbitrary $x, y$ the solution for $Z_A(x,y)$ is

$$Z_A(x,y) = (i)^A C_A^{(x)}(y/2i) \qquad (71)$$

Some examples are

$$Z_1(x,y) = xy$$
$$Z_2(x,y) = x + \frac{1}{2!}x(x+1)y^2 = x + \frac{1}{2!}[x]_2 y^2 \qquad (72)$$
$$Z_3(x,y) = x(x+1)y + \frac{1}{3!}x(x+1)(x+2)y^2 = [x]_2 y + \frac{1}{3!}[x]_3 y^3$$

To further illustrate a point, the following modified generating function is considered:
$1/(1-u-z^2u^2) = \sum_{A=0}^{\infty} Z_A(x,z)u^A$. The solution for $Z_A(x,z)$ is: $Z_A(x,z) = i^A z^A C_A^{(x)}(1/2iz)$.
The result follows from a change of variable $v = zu$. When $z \to 0$ the generating function is that of the permutation case and

$$Z_A(x,z) = i^A z^A C_A^{(x)}(1/2iz)\big|_{z \to 0} \to \frac{x(x+1)(x+2)....(x+A-1)}{A!} \qquad (73)$$

Once $x_k$ is specified, the $Z_A(\vec{x})$ can be calculated and the partition average quantities such as $<n_k>$ follow. For Fibonaaci polynomials generated by $1/(1-yu-u^2)$, the $Z_A(\vec{x}) \to F_A(y) = (i)^A U_A(y/2i)$, with $x = 1$ in $x_k$, and therefore

$$<n_k> = x_k \frac{Z_{A-k}(\vec{x})}{Z_A(\vec{x})} = \frac{1}{k}(g_+^k(y) + g_-^k(y)) \frac{(i)^{A-k} U_{A-k}(y/2i)}{(i)^A U_A(y/2i)} = \frac{1}{k} L_k(y) \frac{F_{A-k}(y)}{F_A(y)} \qquad (74)$$

At $y = 1$

$$<n_k> = \frac{1}{k} L_k \frac{F_{A-k}}{F_A} \to \frac{1}{k} \qquad (75)$$

as obtained before. Even for $y \neq 1$, $<n_k> \sim 1/k$ as shown in Fig.6. In fact, the larger the value of $y$, the greater the range over which the hyperbolic power law is obeyed.

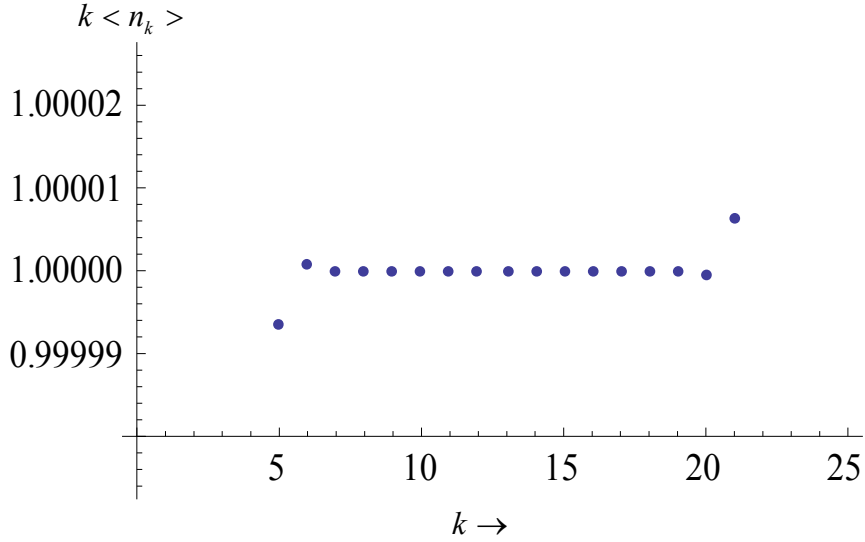

Fig.7. Power law behavior based on Fibonacci/Lucas polynomial model. The value of $A$ is 25 and the results are for $x = 1, y = 3$. The horizontal line at 1 for $k<n_k>$ means that $<n_k> = 1/k$. This hyperbolic power behavior at $x = 1, y = 3$ has the same features as the hyperbolic power behavior at $x = 1, y = 1$.

The constraint $A = \Sigma k n_k$ leads to $AF_A(y) = \Sigma_{k=1}^{A} L_k(y) \cdot F_{A-k}(y)$. The Gegenbauer polynomial $C_A^{(x)}(y)$ is related to the hypergeometric function $_2F_1[a,b,c;z]$: $C_A^{(x)}(y) = ([2x]_A / A!) \cdot _2F_1[-A, 2x+A, x+1/2; (1-y)/2]$. The $_2F_1[-A, b, c; z]$ is a finite polynomial of order $A$ since $[-A]_J = (-A)(-A+1)(-A+2)...(-A+J-1) = 0$ which occurs at $J = A+1$.

II.F. Generalized model and application to an approximate model of Bose-Einstein condensation; connection with links.
F.1 General results
The hypergeometric function $_2F_1[a,b,c,z]$ appeared in the grand canonical generating function for the 4 cases considered in II.B-II.E with the same value for $b, c$, i.e, $b = 1, c = 2$. The cases differ in the choice of $a$ as discussed. The model in II.D has a variable $a$ and the results in II.B and II.C are special cases of II.D with $a = 1$ and $a = 1/2$, respectively. A more general case has variable $a, b, c$. Namely, the grand canonical partition can be taken as

$$Z_{gc} = \exp[xtu(_2F_1(a,b,c,tu))] = \exp\left[xtu\sum_{n=0}^{\infty}\frac{[a]_n[b]_n}{[c]_n n!}t^n u^n\right] = \sum_{A=0}^{\infty} Z_A u^A \qquad (76)$$

Thus

$$x_k = \frac{[a]_{k-1}[b]_{k-1}}{[c]_{k-1}(k-1)!} xt^k \to \frac{1}{k^{1+c-a-b}} \frac{\Gamma[c]}{\Gamma[a]\Gamma[b]} xt^k \qquad (77)$$

where the last part of this equation is the behavior of $x_k$ for large $k$. The associated canonical partition function can be obtained from the recurrence relation:

$$Z_A(x,a,b,c) = \frac{1}{A}\sum_{k=1}^{A} k \frac{[a]_{k-1}[b]_{k-1}}{[c]_{k-1}(k-1)!} xZ_{A-k}(x,a,b,c) \qquad (78)$$

The overall $t^A$ factor is omitted. The recurrence relation generates a polynomial of order $A$ in the variable $x$ for $Z_A(x,a,b,c)$.

F.2 Approximate model of Bose-Einstein condensation

For $b=1, c=2$ the value of $1+c-a-b = 2-a$ which is also the range of the exponent $\tau$. Since $a$ is taken to be positive, $\tau < 2$. However, many exponents are above 2 and several have $\tau = 5/2$ such as Bose-Einstein condensation of particles in a three dimensional box. One possible choice which has $\tau = 5/2$ is $a = 1/2, b = 1, c = 3$. The results of this case are briefly discussed here. For large $k$, $x_k \to x(2/\sqrt{\pi})(1/k^{5/2})$. The exact Bose-Einstein model in d-dimensions has $x_k = x/(k \cdot k^{d/2}) + 1/k$ with $x = L^d/\lambda_T^d$, $\lambda_T = h/(2\pi m k_B T)^{1/2}$. Two functions $g_{5/2}, g_{3/2}$ in $d = 3$ dimensions determine the thermodynamic behavior and condensation point. (see Eq. 8.73 and Eq.8.74 in ref.7):

$$xg_{5/2} = x\sum_{k=1}^{\infty} u^k / k^{5/2}, \qquad xg_{3/2} = x\sum_{k=1}^{\infty} u^k / k^{3/2} \qquad (79)$$

The $1/k$ factor in $x_k$ arises from the cycle class weight in $M_2$ while the $1/k^{3/2}$ in $x_k$ for $d = 3$ comes from a random walk picture [1]. A cycle is a closed loop and parallels a random walk returning to its origin. The pure $1/k$ factor comes from the condensate and is important only at very low temperatures. At condensation the fugacity $u \to 1$ and $g_{5/2} \to \varsigma(5/2) = 1.3415$, $g_{3/2} \to \varsigma(3/2) = 2.6124$. In the approximate model, with $x_k = x([1/2]_{k-1}[1]_{k-1}/([3]_{k-1}(k-1)!)) = x([1/2]_{k-1}/([3]_{k-1}))$, the analogue of $g_{5/2}, g_{3/2}$ are simple closed form expressions which are

$$\sum_{k=1}^{\infty} x_k = xu({}_2F_1[1/2,1,3,u]) = \frac{4}{3} x \frac{(1+\sqrt{1-u}-2(1-u))}{1+\sqrt{1-u}} = \frac{4}{3} x \frac{(1-y)(2y+1)}{1+y} \qquad (80)$$

and

$$\sum_{k=1}^{\infty} kx_k = xu({}_2F_1[1/2,2,3,u]) = -\frac{4}{3} x \frac{-2+2\sqrt{1-u}+u\sqrt{1-u}}{u} = \frac{4}{3} x \frac{(1-y)(2+y)}{(1+y)} \qquad (81)$$

The $y = \sqrt{1-u}$. When $u \to 1$, $y \to 0$ and the first sum becomes $4/3\ x$ while the second sum becomes $8/3\ x$. Fig.8 and Fig.9 compares the exact and approximate behaviors of the two sums as a function of $u$. In each part of the figure one line appears because the two results agree to within the thickness of the line.

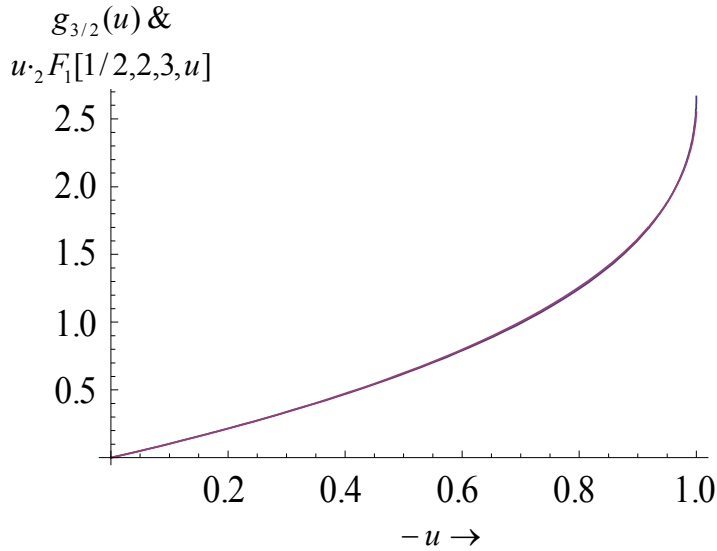

Fig.8. Plot of $g_{3/2}[u]\ \&\ u \cdot (_2F_1[1/2,2,3,u])$ versus $u$. The two agree to within the thickness of the line. The Bose-Einstein condensation endpoints at $u = 1$ are $\varsigma(3/2) = 2.6124$ and $8/3$.

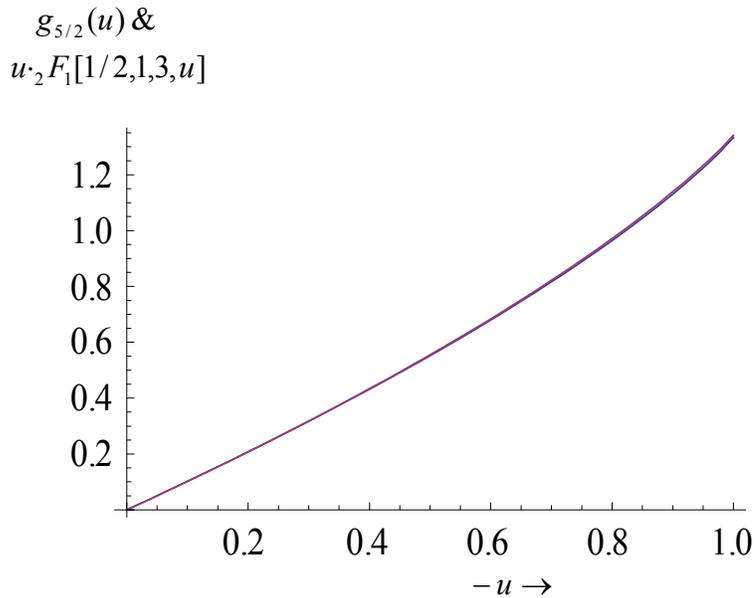

Fig.9. Plot of $g_{5/2}(u)\ \&\ u \cdot F[1/2,1,3;u]$ versus $u$. The two agree to within the thickness of the line. The Bose-Einstein condensation endpoints at $u = 1$ are $\varsigma(3/2) = 1.3415$ and $4/3$.

In the approximate model the mean density $<A>/V$ is connected to the fugacity $u$ by the second sum or:

$$<A> = \sum_{k=1}^{\infty} k x_k = \frac{4}{3}\frac{V}{\lambda_T^3}\frac{(1-y)(2+y)}{(1+y)} \qquad (82)$$

The condensation temperature $T_C$ at fixed density $<A>/V$ is given by $<A>/V = (8/3)/\lambda_{T_C}^3$ while the exact model has $<A>/V = \varsigma(3/2)/\lambda_{T_C}^3$. The last equation can be inverted to find $y$ as a function $D = (3/4)(<A>/V)\lambda_T^3$ by solving a simple quadratic equation $y^2 + (1+D)y + (D-2) = 0$ giving

$$y_{\pm}(D) = \frac{-(1+D) \pm \sqrt{D^2 - 2D + 9}}{2} \qquad (83)$$

Only the $y_+ \equiv y_\pm(D)$ is a physical solution. The fugacity $u$ is $u = 1 - y^2$ for the positive $y$ solution. At $D = (\sqrt{5}-1)/2 = .618034$, the Golden Mean, $y_+(D)$ and $u(D)$ each have the valve $(\sqrt{5}-1)/2$ at this point of intersection.

The pressure is given by

$$\frac{PV}{k_B T} = \sum_{k=1}^{\infty} x_k = xu(_2F_1[1/2,1,3,u]) = \frac{4}{3}x\frac{(1-y)(2y+1)}{1+y} \qquad (84)$$

with $x = V/\lambda_T^3$. The pressure as a function of the density $\rho = <A>/V$ and temperature $T$ can be obtained from the inverted equation of $y$ as a function of $(<A>/V)\lambda_T^3$.

A scale invariant power law in the theory arises at the condensation point. In the thermodynamic limit

$$<n_k> = x_k \frac{Z_{A-k}}{Z_A} \rightarrow x_k \frac{\exp(-F_{h,A-k}/k_B T)}{\exp(-F_{h,A}/k_B T)} = x_k \exp(k\mu/k_B T) \rightarrow x_k = x\frac{2}{\sqrt{\pi}}\frac{1}{k^{5/2}} \qquad (85)$$

The symbol $F_{h,A}$ is the Helmholtz free energy for a system of size $A$. A subscript $h$ is added so as to distinguish it from the Fibonacci number $F_A$. The $F_h$ satisfies $dF_h = -SdT - PdV + \mu dA$ with $S$ the entropy and $\mu$ the chemical potential. Thus, at the critical point of Bose-Einstein condensation a scale invariant power law distribution of cycle lengths appears with exponent $\tau = 5/2$.

The value of $A$, the size of the system, fluctuates in the grand canonical ensemble. These fluctuations are determined by $<A^2> - <A>^2 = \Sigma_k k^2 x_k u^k$. Since $x_k \sim 1/k^{5/2}$ and $u \rightarrow 1$ at the condensation point, $\Sigma_k k^2 x_k$ diverges. This divergence is reflected in the isothermal compressibility $K_T \equiv -(1/V)(\partial V/\partial P)|_T$, where $K_T$ is connected to $<A^2>-<A>^2$ by the equation (ref.[7])

$$<A^2>-<A>^2 = <A>^2 k_B T \frac{1}{V} K_T. \qquad (86)$$

F.3 Connection with networks

The cycles, which arise from permutation symmetries associated with Bose-Einstein statistics, can be viewed as links in a network of connections between particles. At the condensation point a power law distribution of connections appears in the network. This scale invariant power law parallels that seen in small world networks. By contrast, in the very high temperature limit of the model only unit cycles are present and no links appear. As noted at the end of the last section, the value of $\tau$ determines which moments of $x_k$ diverge when $u \to 1$. In Bose-Einstein condensation, the value of $\tau$ is determined by the dimension of the system. For particles in a box of dimension $d$, the connection is simply $\tau = 1 + d/2$. The continuum limit of bosons or atoms in a harmonic oscillator or laser trap has $\tau = 1 + d$.

A model connecting Bose-Einstein condensation and complex networks was also developed by Bianconi and Barabasi[58] using a different parallel which assigns an energy to each node and where a link is a particle. The ability of nodes to compete for links is determined by a fitness parameter determined by a distribution. Fig.1 of ref[58] illustrates the mapping between their network model and the Bose gas. They also consider the evolution or growth of the network.

One aspect of the role of growth is easily handled in the present approach. Adding a new member to the system, which is done by changing $A$ to $A+1$, changes the partition structure. This evolution is shown in Fig.10. The usefulness of the recurrence relation is apparent since it relates the new partition function to a set of partition functions of smaller sizes:

$$Z_{A+1} = (1x_1 Z_A + 2x_2 Z_{A-1} + 3x_3 Z_{A-2} + ..... + Ax_A Z_1 + (A+1)x_{A+1} Z_0)/(A+1) \qquad (87)$$

The $Z_0 = 1$, $Z_1 = x_1$. In many cases the recurrence relation is a much simpler two term relation. The number of groups with $k$ linked members, i.e. the $<n_k>$ distribution, is then changed to $<n_k> = x_k Z_{(A+1)-k}/Z_{A+1}$ by the addition of a new member. The easiest model to analyze has $x_k = xt^k/k$ which has the simpler connection $Z_{A+1} = (x+A)Z_A t/(A+1)$. Then $<n_k>$ is changed to

$$<n_k> = \frac{x}{k} \frac{(A+1)!}{((A+1)-k)!} \frac{1}{(x+(A+1)-k))....(x+(A+1)-1)} \qquad (88)$$

with $k = 1,2,...,A,A+1$. At the scale invariant self similar point which occurs when $x = 1$, the $<n_k> = 1/k$ for $k = 1,2,...,A,A+1$. The distribution stays rigid and a new group size is added with $A+1$ members and mean number $<n_{A+1}>=1/(A+1)$. Note $k<n_k>=1$ for all $k$, which is a feature that arises from the hyperbolic power law behavior. At $x=1$, $Z_{A+1}=Z_A$ and thus the scaling relation $Z_{A+1}/Z_A=1$ follows.

The Fibonacci/Lucas model has similar features. The partition function $Z_A(x,y)=i^A C_A^{(x)}(y)$ and $Z_{A+1}(x,y)$ satisfies the two term recurrence relation

$$(A+1)Z_{A+1}(x,y) = y(x+A)Z_A(x,y) + (2x+A-1)Z_{A-1}(x,y) \tag{89}$$

At $y=1$, $x=1$, $Z_{A+1}=Z_A+Z_{A-1}$ which is the Fibonacci recurrence relation since $Z_A(1,1)=F_A$. Consequently, the scaling relation for the Fibonacci model is simply $Z_{A+1}/Z_A = \phi = (\sqrt{5}+1)/2$, the golden mean, for large $A$. At $x=1$ and arbitrary $y$, the $A!$ will no longer be present in $Z_A$ since $A+1$ is eliminated on both sides of the recurrence equation. When $A$ is changed to $A+1$ the distribution of $<n_k>$ at $x=1$, $y=1$ is changed to

$$<n_k> = \frac{1}{k} L_k \frac{F_{(A+1)-k}}{F_{(A+1)}} \tag{90}$$

In the process a new partition of height $A+1$ is formed in the block picture of Fig.1. The new partition is created by adding a block to the partition in the previous row of height $A$. For $k=A+1$, the

$$<n_{A+1}> = \frac{1}{A+1} L_{A+1} \frac{F_0}{F_{A+1}} \tag{91}$$

The $F_0=1$, $L_{A+1} \to \phi^{A+1}$ and $F_{A+1} \to \phi^{A+1}\sqrt{5}/\phi$. Using these relations the $<n_{A+1}>$ is

$$<n_{A+1}> = \frac{1}{A+1} \frac{\phi}{\sqrt{5}} \tag{92}$$

In general, the $<n_{A+1}>$ is given by the simple result

$$<n_{A+1}> = x_{A+1} \frac{Z_0}{Z_{A+1}} = \frac{x_{A+1}}{Z_{A+1}} \tag{93}$$

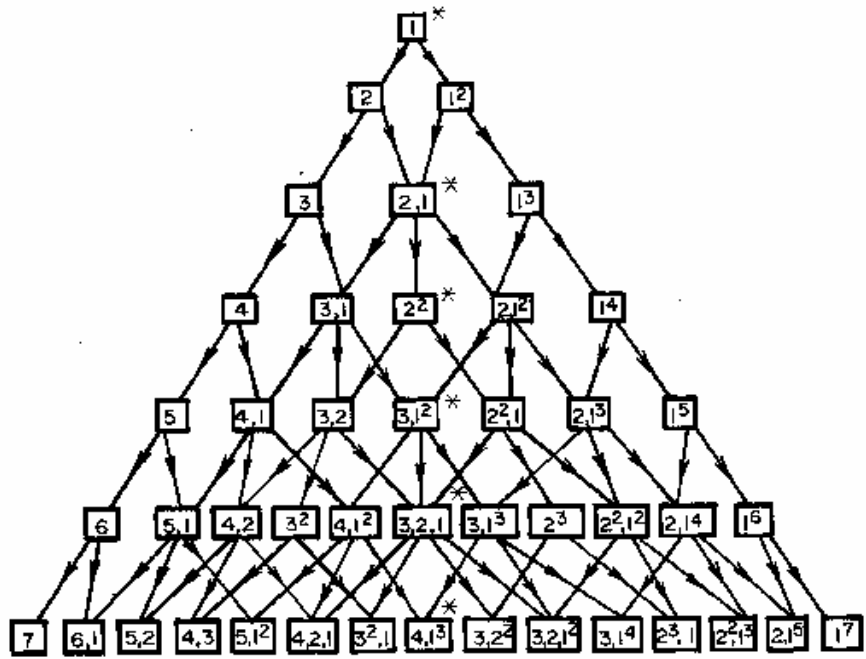

Fig.10. Evolution on a lattice of partitions. Each row has one unit higher $A$ than the previous row. The arrows indicate how each row is connected to the previous row. A block picture of a partition is shown in Fig.1. If an additional block is added to a given block picture in all possible positions, new partitions evolve with the connections shown. For example, starting with the partition 5,1 in row 6 , adding a block: a) to the column 5, generates the partition 6,1, b) adding a block to the column 1, generates the partition 5,2, and finally, c) adding a block alongside of the two columns 5,1 generates $5,1^2$. A common ancestor between any two partitions in a row can be found by tracing the arrows in a backwards direction. In row 7, the partitions $3^2,1$ & $2^2,1^3$ have a common ancestor two rows back in $2^2,1$ in row 5.

Sect.III Conclusions and Summary

This paper explores several exactly soluble models, and in particular contains an extended discussion of a Fibonacci/Lucas based model. In this description a statistical weight is given to each partition of an object into subgroups of varying sizes. The subgroups may represent clusters of varying sizes, cycles in the symmetric group which are important in Bose-Einstein condensation phenomena, group structure or links in networks, the number of copies of a given gene in a sampling of a given size. Summing over all partitions with a fixed constraint on the number of total elements leads to a canonical ensemble partition function of the theory with an associated generating function or grand canonical ensemble partition function. Using these partition functions, partition structure such as the mean distribution of sizes as in a cluster distribution is studied by ensemble averaging. The ensemble averaged distributions in the soluble models considered have a scale invariant power law behavior at a particular critical like point with an associated critical exponent $\tau$. The critical point refers to the value of a tuning parameter labeled $x$ in the approach.

Fibonacci and Lucas polynomials and numbers have played an important role in many areas such as in theories of growth as in biological systems, in discussions of fractals and maps such as in Arnold's cat map, in pure number theory, in geometrical constructions as in Fibonacci spirals, and in aesthetics through their relation to the golden mean. It therefore seemed worthwhile to explore their usefulness and properties when used as a weight function in the theory of partition structures. The following properties were shown in this regard. The Fibonacci/Lucas model has a scale invariant power law behavior at its critical like point with critical index $\tau = 1$. The hyperbolic power law arises in part from the connection of the Fibonacci and Lucas numbers with the Golden mean. The canonical partition function is a Gegenbauer polynomial and the grand canonical partition function or generating function can be written in terms of Gauss hypergeometric functions. The canonical partition function scales as $Z_{A+1}/Z_A \to \phi$, the golden mean. The behavior of the system below the critical point has a $U$ shaped dependence for the cluster size distribution. The behavior of the system above the critical point is also developed. Constraints imposed by the canonical ensemble lead to some number theoretic connections as an incidental consequence of this approach.

The Fibonacci/Lucas model was also compared with other cases previously studied. These were briefly summarized here for the sake of this comparison and for the purposes of a generalization. The generalization involved the Gauss hypergeometric function $_2F_1[a,b,c,u]$ which has an exponent $\tau = 1 + c - a - b$. The permutation model of sect. II.B has $a = 1, b = 1, c = 2$ and therefore $\tau = 1$. Results from the permutation model also appear in population genetics were it is known as the Ewens sampling theory. The Catalan model of sect.II.C which appears in quantum optics when applied to photon count probabilities in lasers, has $a = 1/2, b = 1, c = 2$ and thus $\tau = 3/2$, These two cases are special cases of the Levy index model of sect.II.D which has $b = 1, c = 2$ and $\tau = 2 - a$ with $0 < a \leq 1$. The Fibonacci/Lucas model was related to the permutation case, with results being written as a convolution of the permutation partition functions.

The values of $a, b, c$ in the generalized Gauss hypergeometric function $_2F_1[a,b,c,u]$ can be adjusted to give any desired value of $\tau$, the power law exponent of the theory. For example $a = 1/2, b = 1, c = 3$ and thus $\tau = 5/2$ was shown to be an approximate model of Bose-Einstein condensation for atoms in a three dimensional box. The exact theory and the approximate theory agree to within the thickness of the lines in Fig [8,9]. The zeta function $\varsigma(3/2) = 2.61$ of the exact theory is replaced with 8/3 and $\varsigma(5/2) = 1.34$ with 4/3 in the approximate theory. A connection with the theory of complex networks and links was mentioned. The model is based on a cycle class approach to Bose-Einstein condensation. The cycles, which arise from permutation symmetries associated with Bose-Einstein statistics, form the links in a complex network of connections between particles.

Acknowledgements. This work was supported by the Department of Energy under grant

number DE-FG02-96ER-40987